\documentstyle[12pt,twoside,graphicx]{article}
\begin{document}

\title{Gaussian superpositions in scalar-tensor quantum cosmological models}

\author{R. Colistete Jr.,\thanks{e-mail address: {\tt coliste@ccr.jussieu.fr}}
\\Laboratoire de Gravitation et Cosmologie Relativistes, Universit\'e Pierre\\
et Marie Curie, Tour 22, 4\`eme \'etage, Bo\^{\i}te 142, 4 place Jussieu,\\
75252 Paris Cedex 05, France
\and
J. C. Fabris\thanks{e-mail address: {\tt fabris@cce.ufes.br}}\\
Departamento de F\'{\i}sica, Universidade Federal do Esp\'{\i}rito Santo,\\
29060-900 -- Vit\'oria, Esp\'{\i}rito Santo -- Brazil
\and
and N. Pinto-Neto\thanks{e-mail address: {\tt nelsonpn@lafex.cbpf.br}}\\
Centro Brasileiro de Pesquisas F\'{\i}sicas -- Lafex, Rua Dr.\ Xavier \\
Sigaud 150, Urca 22290-180 -- Rio de Janeiro, RJ -- Brazil}

\date{\today}

\maketitle

\begin{abstract}

A free scalar field minimally coupled to gravity model is quantized and
the Wheeler-DeWitt equation in minisuperspace is solved analytically,
exhibiting positive and negative frequency modes. The analysis is
performed for positive, negative and zero values of the curvature of the
spatial section. Gaussian superpositions of the modes are constructed,
and the quantum bohmian trajectories are determined in the framework of
the Bohm-de Broglie interpretation of quantum cosmology. Oscillating
universes appear in all cases, but with a characteristic scale of the
order of the Planck scale. Bouncing regular solutions emerge for the
flat curvature case. They contract classically from infinity until a
minimum size, where quantum effects become important acting as repulsive
forces avoiding the singularity and creating an inflationary phase,
expanding afterwards to an infinite size, approaching the clas\-sic\-al
expansion as long as the scale factor increases. These are non-singular
solutions which are viable models to describe the early Universe.

\vspace{0.7cm}

PACS number(s): 98.80.Hw, 04.20.Cv, 04.60.Kz

\end{abstract}

\section{Introduction}

The existence of an initial singularity is one of the major drawbacks of
classical cosmology. In spite of the fact that the standard cosmological
model, based in the classical general relativity theory, has been
successfully tested until the nucleosynthesis era (around $t \sim 1 s$),
the extrapolation of this model to higher energies leads to a breakdown
of the geometry in a finite cosmic time. This breakdown of the geometry
may indicate that the classical theory must be replaced by a quantum
theory of gravitation: quantum effects may avoid the presence of the
singularity, leading to a complete regular cosmological model.

The quantization of gravity is plagued with many conceptual and
technical problems, and when it is applied to the whole universe new
issues appear. In the Dirac quantization approach, a functional
equation for the wave function of the Universe is obtained, the
Wheeler-DeWitt equation \cite{wdw}, which is the basic equation of
quantum cosmology. It is formulated in the so-called superspace, the
space of all possible three-dimensional spatial geometries. It is very
hard to find exact solutions of the full Wheeler-DeWitt equation, but
solutions may be found in minisuperspaces where all but a finite number
of degrees of freedom are frozen.

Among the fundamental questions that come from the quantization of the
universe as a whole, one of the most important concerns the
interpretation of the wave function coming from the Wheeler-DeWitt
equation. In order to extract predictions from the wave function of the
Universe, the Bohm-de Broglie ontological interpretation of quantum
mechanics \cite{bohm,holland} has been proposed
\cite{nelson,santini,bola}, since it avoids many conceptual difficulties
that follow from the application of the standard Copenhagen
interpretation to an unique system that contains everything. In
opposition to the latter one, the ontological interpretation does not
need a classical domain outside the quantized system to generate the
physical facts out of potentialities (the facts are there {\it ab
initio}), and hence it can be applied to the universe as a
whole\footnote{Other alternative interpretations can be used in quantum
cosmology like the many worlds interpretation of quantum mechanics
\cite{everett}}.With this interpretation in hands, one can ask if the
quantum scenario predicted by the Wheeler-DeWitt equation is free of
singularities and which type of classical universe emerges from the
quantum phase.

In a preceding work \cite{colistete}, we have applied this proposal to a
free scalar-tensor model with minimal coupling in
Friedmann-Robertson-Walker geometry, which can be obtained from a
non-minimal scalar-tensor theory through a conformal transformation.
Free scalar fields are good candidates to describe the material content
of the early Universe because of their simplicity and because they
represent stiff matter, the type of fluid advocated by Zel'dovich
\cite{zel'dovich} to be relevant at early stages of cosmic evolution.
Only positive curvature spatial sections have been studied. The bohmian
trajectories in configuration space revealed an unexpected scenario:
they behaved as the classical solutions for small values of the scale
factor, but display quantum behaviour when the scale factor is big. As a
consequence, the initial singularity is still present in this quantum
model.

The Wheeler-DeWitt solutions for this scalar-tensor model contain
positive and negative frequency modes, the first leading to an expanding
universe, and the second to a contracting universe, near the
singularity. Inspired by this observation, we constructed in
\cite{colistete} some particular superpositions mixing negative and
positive models. In this way, we found non-singular quantum solutions
which were, however, of planckian size and hence they could not be a
model for our real Universe.

The aim of the present work is to explore further the possibilities of
the minisuperspace model of Reference \cite{colistete}. First, we will
not restrict ourselves to positive curvature spatial sections and
second, we will explore more suitable superpositions of negative and
positive modes, namely, the gaussian superposition. For the case the
spatial section is flat, it is possible to solve analytically the
expressions for the phase of the wave function, and to reduce the
equations for the bohmian trajectories to a dynamical system. The
critical points are calculated, and they are identified as center or
nodes points. This leads to the existence of three kind of scenarios:
periodic solutions representing oscillating universes; bouncing
universes; models with a big-bang followed by a big-crunch. The
bouncing universes contract classically from infinity until a minimum
size, where quantum effects become important acting as repulsive forces
avoiding the singularity, expanding afterwards to an infinite size,
approaching the classical expansion as long as the scale factor
increases. These are non-singular solutions which are viable models to
describe the Universe we live in. For closed and open spatial sections,
all calculations must be performed numerically, and the trajectories
obtained in the configuration space reveal again the presence of
oscillating universes besides those with a big-bang followed by a
big-crunch. In all three cases, the oscillating universes have a
characteristic scale of the order of the Planck length, except for very
special gaussians in the case of zero saptial curvature. Hence, the
most interesting scenarios emerge from the flat case, where we have
succeeded to obtain a viable non-singular model.

The article is organized as follows. In section 2, we describe the
classical model and the corresponding Wheeler-DeWitt equation in the
minisuperspace. Section 3 is devoted to the study of the gaussian
superposition of the quantum solutions found before, and their
corresponding analysis. In section 4 we present our conclusions.

\section{The classical and quantum minisuperspace models}

Let us take the lagrangian
\begin{equation}
\label{lg1}
{\it L} = \sqrt{-g}e^{-\phi}\biggr(R - w \phi_{;\rho}\phi^{;\rho}\biggl)
\quad.
\end{equation}
For $w=-1$ we have effective string theory without the Kalb-Ramond
field. For $w=-3/2$ we have a conformally coupled scalar field.
Performing the conformal transformation $g_{\mu\nu} = e^{\phi}{\bar%
g}_{\mu\nu}$ we obtain the following lagrangian:
\begin{equation}
\label{lg2}
{\it L} = \sqrt{-g}\biggr[R - \biggr(\omega + \frac{3}{2}\biggl)\phi_{;\rho}
\phi^{;\rho}\biggl] \quad,
\end{equation}
where the bars have been omitted. We will define $C_w \equiv (\omega +%
\frac{3}{2})$, which we will consider, from now on, to be strictly
positive in order not to violate any of the energy conditions, at least
classically. 

We will consider the Robertson-Walker metric
\begin{equation}
\label{m}
ds^2 = -N^2 {\rm d}t^2 + \frac{{a(t)}^2}{1 + \frac{\epsilon}{4}r^2}[{\rm
d}r^2 + r^2({\rm d} \theta ^2 + \sin ^2 (\theta) {\rm d} \varphi ^2)] \quad,
\end{equation}
where the spatial curvature $\epsilon$ takes the values $0$, $1$,$-1$.
Inserting this line element into the lagrangian (\ref{lg2}), and using the
units where $\hbar = c = 1$, we obtain the following action:
\begin{equation}
\label{lq}
S = \frac{3 V}{4 \pi l_p^2}\int\frac{Na^3}{2}\biggr(\frac{-{\dot a}^2}{N^2a^2}
+ C_w\frac{{\dot\phi}^2}{6N^2}+ \frac{\epsilon}{a^2}\biggl){\rm d}t \quad,
\end{equation}
where $V$ is the total volume divided by $a^3$ of the spacelike
hypersurfaces, which are supposed to be closed, and $l_p$ is the Planck
length. $V$ depends on the value of $\epsilon$ and on the topology of
the hypersurfaces. For $\epsilon =0$, $V$ can have any value because the
fundamental polyhedra of $\epsilon =0$ hypersurfaces can have arbitrary
size (see Ref. \cite{top}). In the case of $\epsilon = 1$ and topology
$S^3$, $V=2\pi ^2$. Defining $\beta ^2 = \frac{4 \pi l_p^2}{3V}$, ${\bar%
\phi} \equiv \sqrt{\frac{C_w}{6}} \phi$, and omitting again the bars,
the hamiltonian reads
\begin{equation}
\label{hamaphi}
H = N\biggr(-\beta ^2\frac{p_a^2}{2a} + \beta ^2 \frac{p_{\phi}^2}
{2 a^3} - \epsilon\frac{a}{2\beta ^2}\biggl) \quad.
\end{equation}
where 
\begin{eqnarray}
p_a &=& -\frac{a\dot a}{\beta ^2N} \quad, \\
p_\phi &=& \frac{a^3\dot\phi}{\beta ^2N} \quad.
\end{eqnarray}
Usually, the scale factor has dimensions of length because we use
angular coordinates in closed spaces. Hence we will define a
dimensionless scale factor $\tilde{a} \equiv a/\beta$. In that case the
hamiltonian becomes, omitting the tilde:
\begin{equation}
H = \frac{N}{2\beta}\biggr(-\frac{p_a^2}{a} + \frac{p_{\phi}^2}
{a^3} - \epsilon a\biggl) \quad.
\end{equation}
As $\beta$ appears as an overall multiplicative constant in the
hamiltonian, we can set it equal to one without any loss of generality,
keeping in mind that the scale factor which appears in the metric is
$\beta a$, not $a$. We can further simplify the hamiltonian by defining 
$\alpha \equiv \ln(a)$ obtaining
\begin{equation}
\label{hamalphaphi}
H = \frac{N}{2\exp(3\alpha)} \biggr[-p_\alpha ^2 + 
p_{\phi}^2 - \epsilon \exp(4\alpha) \biggl] \quad,
\end{equation}
where
\begin{eqnarray}
\label{palpha}
p_\alpha &=& -\frac{e^{3\alpha}\dot \alpha}{N} \quad, \\
\label{pphi}
p_\phi &=& \frac{e^{3\alpha}\dot\phi}{N} \quad.
\end{eqnarray}

The momentum $p_\phi$ is a constant of motion which we will call ${\bar k}$. 
The classical solutions are, in the gauge $N=1$:
\vspace{0.5cm}

{\bf 1) $\epsilon =0$}

\begin{equation}
\phi = \pm \alpha + c_1 \quad,
\end{equation}
where $c_1$ is an integration constant. In term of cosmic time they read:
\begin{eqnarray}
a &=& e^{\alpha} = 3 {\bar k} t^{1/3} \quad, \\
\phi &=& \frac{\ln (t)}{3} + c_2 \quad.
\end{eqnarray}
The solutions contract or expand forever from a singularity, depending
on the sign of ${\bar k}$, without any inflationary epoch.
\vspace{0.5cm}

{\bf 2) $\epsilon =1$}

\begin{equation}
a = e^{\alpha} = \frac{\bar k}{\cosh(2\phi - c_1)} \quad,
\end{equation}
where $c_1$ is an integration constant, and from the conservation of $p_\phi$
we get
\begin{equation}
{\bar k} = e^{3\alpha}\dot\phi \quad.
\end{equation}
The cosmic time dependence is complicated and we will not write it here.
These solutions describe universes expanding from a singularity till a
maximum size and contracting again to a big crunch. Near the
singularity, these solutions behave as in the flat case. There is no
inflation.
\vspace{0.5cm}

{\bf 3) $\epsilon =-1$} 

\begin{equation}
\label{-1}
a = e^{\alpha} = \frac{\bar k}{\mid\sinh(2\phi - c_1)\mid} \quad,
\end{equation}
where $c_1$ is an integration constant, and again, from the conservation of 
$p_\phi$ we get
\begin{equation}
{\bar k} = e^{3\alpha}\dot\phi \quad.
\end{equation}
As before, the cosmic time dependence is complicated and we will not
write it here. These solutions describe universes contracting forever to
or expanding forever from a singularity. Near the singularity, these
solutions behave as in the flat case. There is no inflation\footnote{In
the case $\epsilon =-1$ there are classical solutions with $C_w < 0$.
Qualitatively, they represent universes contracting from an infinite 
to a minimum size and then expanding again to infinity}.
\vspace{0.5cm}

Let us now quantize the model. The Wheeler-DeWitt equation is obtained
through the Dirac quantization procedure where the wave function must be
annihilated by the operator version of the constraint in Eq.
(\ref{hamalphaphi}). With the choice of factor ordering which makes it
covariant through field redefinitions, it reads 
\begin{equation}
\label{wdw}
- \frac{\partial ^2\Psi}{\partial \alpha ^2} +  \frac{\partial
^2\Psi}{\partial \phi ^2} + \epsilon e^{4\alpha}\Psi = 0 \quad.
\end{equation}
Employing the separation of variables method, we obtain the general solution
\begin{equation}
\Psi(\alpha,\phi) = \int{F(k)A_k(\alpha)B_k(\phi)dk} \quad,
\end{equation}
where $k$ is a separation constant, 
\begin{equation} 
B_k(\phi) = b_1\exp(ik\phi) + b_2\exp(-ik\phi) \quad,
\end{equation}
and for $\epsilon =0$
\begin{equation} 
A_k(\alpha) = a_1\exp(ik\alpha) + a_2\exp(-ik\alpha) \quad,
\end{equation}
for $\epsilon =1$
\begin{equation} 
A_k(\alpha) = a_1 I_{ik/2}(e^{2\alpha} /2) + a_2 K_{ik/2}(e^{2\alpha} /2)
\quad,
\end{equation}
and for $\epsilon =-1$
\begin{equation} 
A_k(\alpha) = a_1 J_{ik/2}(e^{2\alpha} /2) + a_2 N_{ik/2}(e^{2\alpha} /2)
\quad.
\end{equation}
The functions $J, N, I, K$ are Bessel and modified Bessel functions of
first and second kind.

The Bohm-de Broglie interpretation of homogeneous minisuperspace models 
goes as follows: in general, the minisuperspace Wheeler-De Witt equation is
\begin{equation} 
\label{bsc}
{\cal H}({\hat{p}}^{\alpha}(t), {\hat{q}}_{\alpha}(t)) \Psi (q) = 0 \quad.
\end{equation}
Writing $\Psi = R \exp (iS/\hbar)$, and substituting it into (\ref{bsc}),
we obtain the following equation:
\begin{equation}
\label{hoqg}
\frac{1}{2}f_{\alpha\beta}(q_{\mu})\frac{\partial S}{\partial q_{\alpha}}
\frac{\partial S}{\partial q_{\beta}}+ U(q_{\mu}) + Q(q_{\mu}) = 0 \quad,
\end{equation}
where the quantum potential is
\begin{equation}
\label{hqgqp}
Q(q_{\mu}) = -\frac{1}{2R} f_{\alpha\beta}\frac{\partial ^2 R}
{\partial q_{\alpha} \partial q_{\beta}} \quad.
\end{equation}

The Bohm-de Broglie interpretation applied to quantum cosmology states
that the trajectories $q_{\alpha}(t)$ are real, independently of any
observations. Eq. (\ref{hoqg}) is the Hamilton-Jacobi equation for them,
which is the classical one amended with a quantum potential term
(\ref{hqgqp}), responsible for the quantum effects. This suggests to
define
\begin{equation}
\label{h}
p^{\alpha} = \frac{\partial S}{\partial q_{\alpha}} \quad,
\end{equation}
where the momenta are related to the velocities in the usual way:
\begin{equation}
\label{h2}
p^{\alpha} = f^{\alpha\beta}\frac{1}{N}\frac{\partial q_{\beta}}{\partial t}
\quad.
\end{equation}
To obtain the quantum trajectories we have to solve the following
system of first order differential equations, called the guidance relations:
\begin{equation}
\label{h3}
\frac{\partial S(q_{\alpha})}{\partial q_{\alpha}} =
f^{\alpha\beta}\frac{1}{N}\dot{q}_{\beta} \quad.
\end{equation}

In the present case of the hamiltonian (\ref{hamalphaphi}), the quantum
potential (\ref{hqgqp}) becomes
\begin{equation}
Q(\alpha ,\phi )=\frac{e^{3\alpha}}{2R}\biggr[\frac{\partial^{2}R}{%
\partial \alpha^{2}}-\frac{\partial^{2}R}{\partial \phi^{2}}\biggl]\quad ,
\end{equation}
and the guidance relations (\ref{h3}) read
\begin{equation}
\label{guialpha}
\frac{\partial S}{\partial \alpha}=-\frac{e^{3\alpha}\dot{\alpha}}{N}\quad ,
\end{equation}
\begin{equation}
\label{guiphi}
\frac{\partial S}{\partial \phi}=\frac{e^{3\alpha}\dot{\phi}}{N}\quad .
\end{equation}

Eqs. (\ref{h3}) are invariant under time reparametrization. Hence, even
at the quantum level, different choices of $N(t)$ yield the same
spacetime geometry for a given non-classical solution $q_{\alpha}(t)$.
There is no problem of time in the Bohm-de Broglie interpretation of
minisuperspace quantum cosmology\footnote{This is not the case,
however, for the full superspace (see Reference \cite{santini}).}. Let us
then apply this interpretation to our minisuperspace models and choose
the gauge $N = 1$.

\section{Bohm interpretation of gaussian superpositions}

We will now make gaussian superpositions of these solutions and interpret
the results using the Bohm-de Broglie interpretation of quantum mechanics.
We will begin by the case  $\epsilon =0$, which is simpler, and it
is the one to which the others reduce when $\alpha \rightarrow -\infty$.

\subsection{Hypersurfaces with $\epsilon =0$}

This case can be solved analytically. The function $F(k)$ is
\begin{equation} 
F(k) = \exp\biggr[-\frac{(k-d)^2}{\sigma ^2}\biggl] \quad.
\end{equation}
We can study two types of wave function:
\begin{equation}
\Psi _1(\alpha,\phi) = \int{F(k)B_k(\phi)[A_k(\alpha)+
A_{-k}(\alpha)]dk} \quad,
\end{equation}
and
\begin{equation}
\Psi _2(\alpha,\phi) = \int{F(k)A_k(\alpha)[B_k(\phi)+
B_{-k}(\phi)]dk} \quad,
\end{equation}
both with $a_2 = b_2 = 0$. We will restrict ourselves to $\Psi _1$
because it yields the most interesting results. The results coming from
$\Psi _2$ can be obtained from the first by changing $\alpha$ with
$\phi$.

Performing the integration in $k$ we obtain for $\Psi _1$:
\begin{equation}
\label{psi1}
\Psi _1 = \sigma \sqrt{\pi} \biggr\{
\exp\biggr[-\frac{(\alpha + \phi)^2 \sigma ^2}{4}\biggl]
\exp[id(\alpha + \phi)] + 
\exp\biggr[-\frac{(\alpha - \phi)^2 \sigma ^2}{4}\biggl]
\exp[-id(\alpha - \phi)]\biggl\} \quad.
\end{equation}
In order to obtain the bohmian trajectories, we have to calculate the
phase $S$ of the above wave function and substitute it into the
guidance formula (\ref{guialpha}--\ref{guiphi}), working in the gauge
$N=1$. These equations constitute a planar system which can be easily
studied: 
\begin{eqnarray}
\label{dalphadt}
\dot \alpha &=& \frac{\biggl[\phi \sigma ^2 \sin (2d\alpha ) + 
2d\sinh (\sigma ^2 \alpha  \phi) \biggr]}
{\exp (3\alpha)\biggl\{ 2[ \cos (2d\alpha) + 
\cosh (\sigma ^2 \alpha \phi)]\biggr\}} \quad, \\
\label{dphidt}
\dot\phi &=&  \frac{\biggl[-\alpha \sigma ^2 \sin (2d\alpha) +
2d\cos (2d\alpha ) + 2d\cosh (\sigma ^2 \alpha \phi) \biggr]}
{\exp (3\alpha)\biggl\{ 2[ \cos (2d\alpha) + 
\cosh (\sigma ^2 \alpha \phi)]\biggr\}} \quad.
\end{eqnarray}
The line $\alpha = 0$ divides configuration space in two symmetric
regions. The line $\phi = 0$ contains all singular points of this
system, which are nodes and centers. The nodes appear when the
denominator of the above equations, which is proportional to the norm of
the wave function, is zero. No trajectory can pass through these points.
They happen when $\phi = 0$ and $\cos (d\alpha ) = 0$, or $\alpha =%
(2n+1)\pi /2d$, $n$ an integer, with separation $\pi/d$. The center
points appear when the numerators are zero. They are given by $\phi = 0$
and $\alpha = 2 d[{\rm cotan} (d\alpha )]/\sigma ^2$. They are
intercalated with the node points, and their separations cannot exceed
$\pi/d$. As $\mid \alpha \mid \rightarrow \infty$ these points tend to
$n\pi/d$. As one can see from the above system, the classical solutions
($a(t)\propto t^{1/3}$) are recovered when $\mid \alpha \mid \rightarrow
\infty$ or $\mid \phi \mid \rightarrow \infty$, the other being
different from zero. 

A field plot of this planar system is shown in Figure 1, for $\sigma = d
=1$. We can see plenty of different possibilities, depending on the
initial conditions. Near the center points we can have oscillating
universes without singularities and with amplitude of oscillation of
order 1\footnote{As discussed above, these amplitudes can be very large
as long as $d$ becomes very small because the separation of the center
points are of the order of $1/d$}. For negative values of $\alpha$, the
universe arises classically from a singularity but quantum effects become
important forcing it to recollapse to another singularity, recovering
classical behaviour near it. For positive values of $\alpha$, the
universe contracts classically but when $\phi$ and $\alpha$ are small
enough, quantum effects become important creating an inflationary phase
which avoids the singularity. The universe contracts to a minimum size
and after reaching this point it expands forever, recovering the
classical limit when $\alpha$ becomes sufficiently large. These are
models which can represent the early Universe. We can see that for
$\alpha$ negative we have classical limit for small scale factor while
for $\alpha$ positive we have classical limit for big scale factor. 

For the wave function $\Psi _2$, the analysis goes in the same way but
we have to interchange $\alpha$ with $\phi$. In this case we also have
periodic solutions but the others are universes arising classically from
a singularity, experiencing quantum effects in the middle of their
expansion, and recovering their classical behaviour for large values of
$\alpha$. There are no further possibilities.

\begin{figure}[t]
\begin{center}
\includegraphics[scale=0.90, trim=0in 0in 0in 0in]{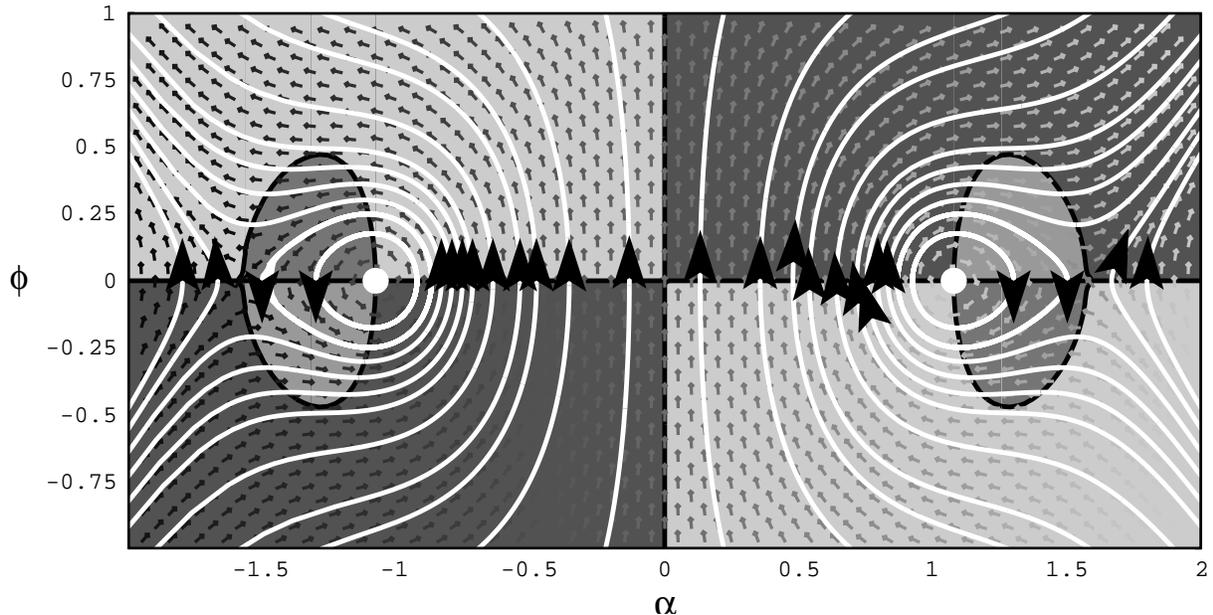}
\caption{
Field plot of the system of planar equations (\ref{dalphadt}-\ref{dphidt})
for $\sigma =d=1$, which uses the Bohm-de Broglie
interpretation with the wave function $\Psi_1$, Equation (\ref{psi1}).
Each arrow of the vector field is shaded according to its true length, black
representing short vectors and white, long ones.
The four shades of gray show the regions where the vector field is pointing
to northeast, northwest, southeast or southwest. The black curves are the 
nullcline curves that separate these regions. The white points are the centers
points whose neighbourhoods have oscillating trajectories. The trajectories
are the white curves with direction arrows.
}
\end{center}
\end{figure}

We will now pass to the cases with curved spatial sections. One can
immediately notice an important difference. The case $\epsilon =0$ has a
symmetry $\alpha \rightarrow -\alpha$ which is present not only in the
Wheeler-DeWitt equation (\ref{wdw}) but also in the solution
(\ref{psi1}). The cases $\epsilon \neq 0$ do not possess this symmetry
(the potential $\epsilon e^{4\alpha}$ in the Wheeler-DeWitt equation
breaks it), and one should not expect to obtain the $\alpha > 0$ part of
the field plots in these cases from the $\alpha < 0$ part through a
reflection, as in the case $\epsilon =0$ (see Figure 1).

\subsection{Hypersurfaces with $\epsilon =1$} 

The Wheeler-DeWitt equation (\ref{wdw}) for $\epsilon =1$, in the case
we neglect the $\partial _{\phi\phi} \Psi$, is analogous to a stationary
Schroedinger equation with $E=0$ and $V=e^{4\alpha}$. Hence, one should
make superpositions involving only the parts of $A_k(\alpha)$ which goes
to zero as $\alpha$ goes to infinity, which are the Bessel functions
$K_{ik/2}(e^{2\alpha /2})$. Consequently, we will take the following
superposition: 
\begin{equation}
\label{psi3}
\Psi_{3} (\alpha,\phi) = \int{\exp\biggr[-\frac{(k-d)^2}{\sigma^2}\biggl]
K_{ik/2}(e^{2\alpha} /2) B_{k}(\phi)dk} \quad.
\end{equation}
\begin{figure}[!b]
\begin{center}
\includegraphics[scale=0.90]{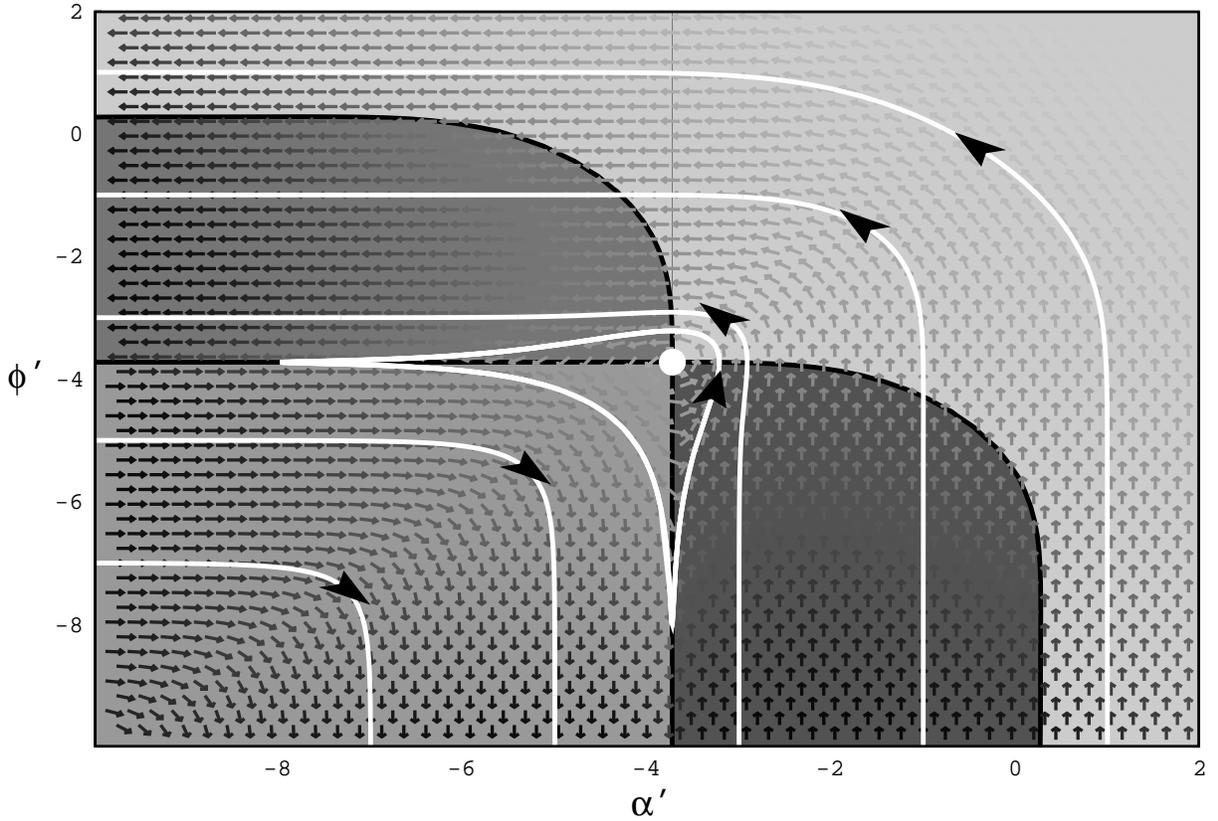}
\caption{
Field plot of the numerical solution of the system of planar equations
(\ref{dalphaprimedt}--\ref{dphiprimedt}) using the Bohm-de Broglie
interpretation with the wave function $\Psi_3$, Equation (\ref{psi3}),
for $\sigma=d=1$. Each arrow of the vector field is shaded according to
its true length, black representing short vectors and white, long ones.
The four shades of gray show the regions where the vector field is
pointing to northeast, northwest, southeast or southwest. The black
curves are the nullcline curves that separate these regions. The white
point is the center point whose neighbourhood has oscillating
trajectories. The trajectories are the white curves with direction
arrows.
}
\end{center}
\end{figure}
The limit $\alpha \rightarrow -\infty$ does not give the preceding
results for $\epsilon =0$ because the Bessel function $K$ reduces in
this limit to
\begin{equation}
K_{ik/2}(e^{2\alpha} /2) \approx \frac{i}{k} \biggr[
\exp[ik(\alpha - \ln (2)] \Gamma (1-\frac{ik}{2}) - 
\exp[-ik(\alpha - \ln (2)] \Gamma (1+\frac{ik}{2})\biggl] \quad,
\end{equation}
and the presence of the Gamma functions spoils their similarity. 

This case must be studied numerically, and the transformation
\begin{eqnarray}
\label{alphaprime}
\alpha ^{\prime } &=&\alpha -\phi \quad ,\\
\label{phiprime}
\phi ^{\prime } &=&\phi +\alpha \quad ,
\end{eqnarray}
eases this task. The guidance relations (\ref{guialpha}--\ref{guiphi}) become
\begin{eqnarray}
\label{dalphaprimedt} 
\dot{\alpha}^{\prime } &=&-2\exp \biggr[\frac{-3(\alpha ^{\prime }+\phi
^{\prime })}{2}\biggl]\frac{\partial S}{\partial \phi ^{\prime }}\quad , \\
\label{dphiprimedt}
\dot{\phi}^{\prime } &=&-2\exp \biggr[\frac{-3(\alpha ^{\prime }+\phi
^{\prime })}{2}\biggl]\frac{\partial S}{\partial \alpha ^{\prime }}\quad ,
\end{eqnarray}
and Figure 2 shows the field plot of this transformed planar system,
using $\sigma=d=1$. There are periodic solutions without singularities
which happen when the bohmian trajectories cross the lines
$(\alpha^{\prime} =-3.73$, $\phi^{\prime} < -3.73)$, or $(\phi^{\prime}%
=-3.73$, $\alpha^{\prime} < -3.73)$, or equivalently $(\alpha = \mid%
\phi \mid -3.73$, $\alpha < -3.73)$. These oscillating trajectories can
reach very negative values of $\alpha$ but their maximum size cannot
exceed $\alpha =0$, or $a \approx l_{pl}$. Another behaviour is related
to the trajectories shown in Figure 2 which are exclusively in the light
gray region. They begin classically from a singularity, expand to a
maximum value of $\alpha$, and then return classically to a singularity.
Concluding, we have two types of trajectories in this case: one which is
periodic due to quantum effects, and the other which exhibit the pattern
of classical behaviour: expansion from a singularity until a maximum size
followed by a contraction to a big crunch. The periodic solutions have
maximum size around $\alpha =0$, or $a \approx l_{pl}$ and they cannot
represent the Universe we live in.

\subsection{Hypersurfaces with $\epsilon =-1$}

In this case we will choose as $A_k (\alpha)$ the combination 
\begin{equation}
\label{AGamma}
A_k (\alpha) = \biggr[\Gamma (1+\frac{ik}{2}) J_{ik/2}(e^{2\alpha} /2) + 
\Gamma (1-\frac{ik}{2}) J_{-ik/2}(e^{2\alpha} /2)\biggl] 
\end{equation}
in order to get rid of the Gamma functions and obtain the preceding
results for $\epsilon =0$ when $\alpha$ is very negative because the
Bessel function $J$ reduces in this limit to
\begin{equation}
J_{ik/2}(e^{2\alpha}/2) \approx 
\frac{\exp[ik(\alpha - \ln (2)]}{\Gamma (1+\frac{ik}{2})} \quad.
\end{equation}
Taking this choice of $A_{k}(\alpha)$, Eq. (\ref{AGamma}), into the gaussian
superposition 
\begin{equation}
\label{psi4}
\Psi_{4} (\alpha,\phi) = \int{\exp\biggr[-\frac{(k-d)^2}{\sigma^2}\biggl]
A_{k}(\alpha) B_{k}(\phi)dk} \quad,
\end{equation}
the numerical calculations with respect to the $\epsilon=-1$ case show
that the behaviour for very negative values of $\alpha$ is similar to
the $\epsilon=0$ case, as one can see by comparing Figure 3 with Figure
1. As $\alpha$ increases the regions with oscillating universes are
squeezed and their separation decrease monotonically. Like the
$\epsilon=0$ case, there are periodic solutions without singularities
and with amplitude of oscillation of order $1$. The other behaviour
is described by trajectories that arise classically from a singularity,
experiment a quantum halt at some maximum value of the scale factor, and
then classically contracts to a big-crunch, contrary to the classical
solutions of Eq. (\ref{-1}) which contract forever to or expand forever
from a singularity. 

\begin{figure}[!t]
\begin{center}
\includegraphics[scale=0.90]{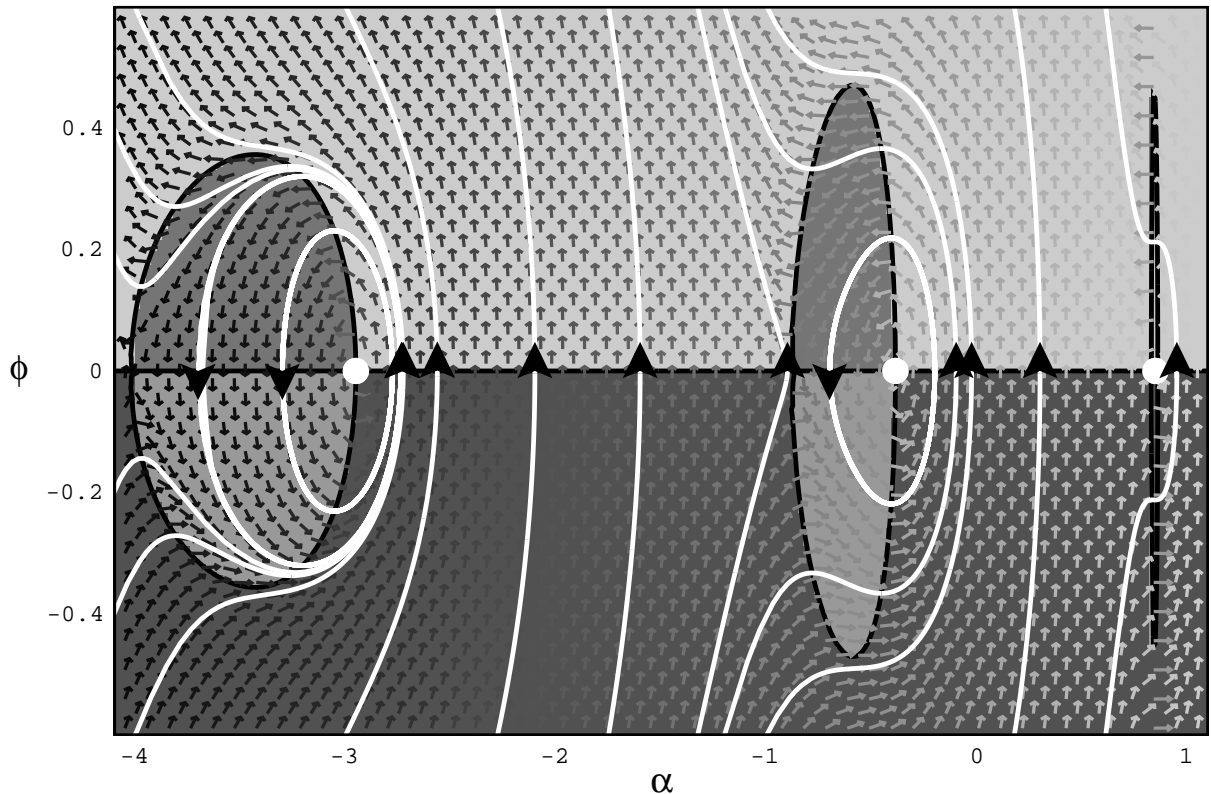}
\caption{
Field plot of the numerical solution of the system of planar equations
(\ref{guialpha}--\ref{guiphi}) using the Bohm-de Broglie interpretation
with the wave function $\Psi_4$, Equation (\ref{psi4}), for $\sigma=d=1$.
Each arrow of the vector field is shaded according to its true length,
black representing short vectors and white, long ones. The four shades
of gray show the regions where the vector field is pointing to
northeast, northwest, southeast or southwest. The black curves are the
nullcline curves that separate these regions. The white points are the
centers points whose neighbourhoods have oscillating trajectories. The
trajectories are the white curves with direction arrows.
}
\end{center}
\end{figure}

\section{Conclusion}

The quantization of a scalar-tensor model in the minisuperspace leads to
a separable partial differential equation, admitting analytical
solutions, with positive and negative frequencies. In this work, we have
studied gaussian superpositions of these different modes and the
corresponding bohmian trajectories. Such analysis was performed for
zero, positive and negative curvature spatial sections, which are
considered to be compact. The bohmian trajectories in configuration
space were calculated numerically, excepted for the flat case, where it
is possible to reduce the equations for the bohmian trajectories to a
two dimensional dynamical system.

The comparison of the trajectories in the configuration space of the
variables $a$ and $\phi$, which are the dynamical degrees of freedom of
the minisuperspace, with the classical ones, allows one to identify the
classical and quantum phases for the scalar-tensor cosmological models.
For all three different values of the curvature of the spatial sections,
the configuration space of the quantum solutions displays oscillating
universes. However, these oscillating universes remain at the Planck
scale and they cannot be considered as candidates for the description of
the early Universe (they are more like baby universes), except for the
unnatural choice $|d|<<1$ in the $\epsilon =0$ case. There are also
trajectories which correspond to universes which begin and end in
singular states. Only for the flat case it is possible to have bouncing
models.

In the bouncing models of the flat spatial section case, the scale
factor has an infinite initial and final values, near which it behaves
classically. As it approaches the singularity, the repulsive quantum
effects lead to the bounce, avoiding the singularity. Such a scenario
can be a candidate for the description of our early Universe, since it
is free from the initial singularity and behaves classically in the
asymptotic limit of large values for the scale factor. It is worth to
note that this classical asymptotic limit corresponds to the stiff
matter Friedmann universe which, according to Zel'dovich
\cite{zel'dovich}, is the most promising one to describe the very early
Universe.

The free scalar field model considered in this paper, on the other hand,
can be connected to a non-minimal coupled scalar field, with a coupling
parameter $\omega$, like in the Brans-Dicke theory, by a conformal
transformation. In \cite{velasco} a quantum analysis of these
non-minimal models was performed, and it was shown that non-singular
scenarios can be obtained when the parameter $\omega$ is negative. In
fact, all quantum analysis performed here can be connected with a
similar analysis in the non-minimal case through a conformal
transformation, namely $\alpha _{NMC} = \alpha +\phi$ and $\phi _{NMC} =
\phi$. One can verify that when the minimal model displays
singularities, it is possible to have non-singular solutions in the
corresponding non-minimal case; but the non-singular solutions in the
minimal case must also be non-singular in the corresponding non-minimal
models.

An important generalization of the model studied here would be to
consider self-interacting scalar fields. It was shown in \cite{sergio}
that to each perfect fluid barotropic equation of state, it is possible
to construct a self-interacting scalar field model leading to the same
classical description in minisuperspace. We may argue if this
correspondence remains at the quantum level. Note that in Ref.
\cite{bola} we have obtained bouncing universes for radiation fields
with $\epsilon =0$ and $\epsilon =-1$ with the same qualitative
behaviour as the bouncing universes found here. They are also viable
models for the early Universe. One should investigate if a scalar field
model with a potential corresponding to the radiation fluid would give
similar results. 

\section*{ACKNOWLEDGMENTS}

We would like to thank Andrew Sornborger and the Cosmology Group of CBPF
for many useful discussions, and CNPq and CAPES of Brazil for financial
support. One of us (NPN) would like to thank the Laboratoire de
Gravitation et Cosmologie Relativistes of Universit\'e Pierre et Marie
Curie, and Fermilab, where part of this work has been done, for
financial aid and hospitality.

\end{document}